\documentstyle[aps,prl]{revtex}
\begin{document}
\draft
\title{Fast Zonal Field Dynamo in Collisionless Kinetic Alfven Wave Turbulence}
\author{I. Gruzinov, A. Das\thanks{Institute for Plasma Research, Bhat, Gandhinagar, 382428 India}, P.H. Diamond} 
\address{University of California, San Diego, La Jolla, CA 92093-0319 USA}
\author{A. Smolyakov}
\address{University of Saskatoon, Saskatchewan, SK S7N 5E2, Canada}

\maketitle

\begin{abstract}
The possibility of fast dynamo action by collisionless kinetic Alfven Wave turbulence is demonstrated. The irreversibility necessary to lock in the generated field is provided by electron Landau damping, so the induced electric field does not vanish with resistivity.  Mechanisms for self-regulation of the system and the relation of these results to the theory of alpha quenching are discussed. The dynamo-generated fields have symmetry like to that of zonal flows, and thus are termed zonal fields.\\
\end{abstract}

pacs{52.35.Vd, 9.25.Cw, 52.35.Mw}\\

The dynamo problem, i.e. the problem of understanding the origin and generation mechanism for astrophysical and geophysical magnetic fields, remains one of the major unsolved problems in classical physics \cite{ref1}.  Until recently, the kinematic quasilinear theory of the alpha effect provided an attractive framework for formulating a solution to this problem \cite{ref2}.  In this approach, quasilinear iteration was used to relate the turbulence-induced mean EMF, i.e.  $\langle{\underline{\widetilde{v}}} \times {\underline{\widetilde{B}}}\rangle$ to $\alpha \langle\underline{B} \rangle - \beta \langle \underline{J}\rangle$, where $\alpha$, the dynamo drive term is in turn proportional to the helicity of the fluid turbulence, i.e. $\langle{\widetilde{\underline{v}}} \cdot {\widetilde{\underline{\omega}}}\rangle$.  Recent computational and theoretical research has, however, revealed that the kinematic theory is invalid, since the small scale magnetic field grows on fast, inertial range time scales, and so cannot be neglected when considering the evolution of the mean-field \cite{ref3}.  The `bottom line' of this research is that $\alpha_{K}$, the kinematic alpha coefficient discussed above, is substantially reduced by self-consistent small scale field effects, so that the net alpha effect is \cite{ref4}

\begin{equation}
{\alpha \equiv} {{{\alpha_{K}}} \over {\lbrack 1 + {R_{m}} {v_{Ao}^{2}} / {\langle {\widetilde{v}^{2} \rangle} \rbrack}}}\label{eq1}.
\end{equation}

\noindent
Here $v_{Ao} = {\langle \underline{B} \rangle / \sqrt{4\pi\rho_0}}$ is the Alfven speed in the mean field, $\langle \widetilde{v}^{2} \rangle$ is the mean square turbulence velocity, $R_{m}$ is the magnetic Reynolds number and $\alpha_{K} = \sum\limits_{\underline{k}} \langle{\underline{\widetilde{v}}} \cdot {\underline{\widetilde{\omega}}}{\rangle_{\underline{k}} \tau_{c},_{\underline{k}}}$.  $\tau_{c},_{\underline{k}}$ is an eddy self-correlation time.  Eqn. (\ref{eq1}) states that the alpha effect is quenched for $v_{Ao} > ({1 / \sqrt{R_{m}}}) \widetilde{v}_{rms}$, i.e. for very small $\langle{\underline{B}}\rangle$, since $R_{m} >> 1$ in most relevant applications.  Note that the above expression for $\alpha$ may be re-written as $\alpha = {\alpha_{K}} {\eta / {\lbrack\eta + \tau_{c} v_{Ao}^{2}}} \rbrack$, where $\tau_{c}$ is the integral scale turbulence correlation time.  This expression suggests that one way of interpreting the alpha `quench' is that despite naive expectations of turbulent mixing, etc., the collisional resistivity ultimately controls the rate at which `stretched and twisted' fields are folded to increase $\langle {\underline{B}} \rangle$.  This constraint necessarily precludes the possibility of a fast dynamo, i.e. one which operates on time scales independent of $\eta$.  Another possible interpretation of the quench is that of `Alfvenization', whereby the growing $\langle{\underline{B}} \rangle$ converts eddys into Alfven waves (which are intrinsically non-kinematic), thus initiating the quench.  Indeed, with this in mind, it is interesting to note that the EMF for an individual, circularly polarized visco-resistive MHD Alfven wave is $<\widetilde{\underline{v}} \times \widetilde{\underline{B}}> = (\eta / 2B_{o}) (1 - v/\eta) k^{2}k_{z}\widetilde{A}^{2}$\cite{ref5}.  Apart from the `cross-helicity' factor $(1 - {\nu / \eta})$ and a somewhat different expression for the helicity spectrum, this formula is identical to that given by Eqn. (\ref{eq1}), thus re-inforcing the notion that Alfvenization is at work in the alpha quench.

In this paper, we report on the theory of a fast, collisionless dynamo.  In view of the above discussion of Alfvenization, we proceed to directly consider a dynamo driven by Kinetic Shear Alfven Wave (KSAW) turbulence \cite{ref6}.  Here, the dynamo instability mechanism is that of a modulational instability of the KSAW spectrum.  In this case, the dynamically generated field is orthogonal to an externally prescribed mean field $\underline{B}_{o} = B_{o}{\widehat{z}}$, along which the KSAW's propagate \cite{ref7}.  Collisionless dissipation, via Landau damping, locks in the dynamo-generated field at a rate independent of $\eta$, thus facilitating fast dynamo action.  Since, assuming radially inhomogeneous turbulence, the dynamo-generated fields have azimutal symmetry and are thus analogous to zonal flows \cite{ref8}, we hereafter refer to them as zonal fields.

As the current is carried by the electrons in KSAW turbulence, we start from the drift-kinetic equation

\begin{equation}
{\partial f \over \partial t} + {v_{\parallel}}{\nabla_{\parallel}}f-{{c}\over{B_{o}}}{\underline{\nabla}}\phi\times{\widehat{z}}\cdot{\underline{\nabla}}f-{{{\vert} e {\vert}}\over{m_{e}}}{E_{\parallel}} {{\partial f} \over {\partial v_{\parallel}}} = c (f)\label{eq2}.
\end{equation}

\noindent
Here $E_{\parallel} = - \nabla_{\parallel}\phi - ({1 / c}) {{\partial A} / {\partial t}}$, where $\phi$ is the electrostatic potential, $A$ is the $\widehat{z}$ component of the vector potential, and $\nabla_{\parallel} = {{\partial} / {\partial z}} + \underline{\nabla}(A / B_{o}) \times {\widehat{z}} \cdot \nabla$.  Neglecting electron inertia, averaging Eqn.(\ref{eq2}), and using Ampere's Law then yields the mean field Ohm's Law for collisionless plasma:

\begin{equation}
{1 \over c} {{\partial \langle A \rangle} \over {\partial t}} - {\langle {{\widetilde{E}_{z}} {\widetilde{n} \over n_{o}}} \rangle} + {\partial \Gamma \over \partial r} = {\eta \over c} {\nabla_{\perp}^{2}} {\langle A \rangle}\label{eq3}.
\end{equation}

\noindent
Terms $\langle {\widetilde{E}}_{z} {\widetilde{n} / n_{o}} \rangle$ and ${\partial\Gamma}/{\partial r}$ refer to parallel electron acceleration and the spatial transport of parallel current, respectively.  The flux $\Gamma$ is given by ${(1 / {\Omega_{e}})} \int d {v_{\parallel}} {v_{\parallel}} \langle({\nabla_{\theta}}(\phi - ({v_{\parallel}} / c) \widetilde{A}_{\parallel})) \widetilde{f} \rangle$.  Proceeding as in quasilinear theory, we can use the coherent response $\widetilde{f}_{k}$:

\begin{mathletters}
\begin{equation}
{\widetilde{f}_{k}} = {e \over T_{e}} {(\phi - \psi)_{k}} {f_{o} -} {e \omega_{\underline{k}} \over T_{e}} {{(\phi - \psi)_{k} f_{o}} \over (\omega_{k} - k_{\parallel} v_{z})}\label{eq4a},
\end{equation}

\noindent
to calculate $\langle{\widetilde{E}_{z}}{\widetilde{n} / n_{o}}\rangle$ and $\Gamma$, so that

\begin{equation}
{\langle {{\widetilde{E}_{\parallel}} {\widetilde{n} \over n_{o}}} \rangle} = - {\pi T_{e} \over e} {\sum_{k}} {k_{\parallel}} {\left| {{e(\phi-\psi)_{k}} \over {T_{e}}} \right|^{2}} {({{\omega} \over {\vert k_{\parallel} \vert}})} f_{o} {\vert_{\omega / k_{\parallel}}}\label{eq4b}
\end{equation}

\begin{equation}
\underline{\Gamma}= {\pi T_{e} \over e \Omega_{e}} \sum_{k} \underline{k} \times \widehat{z} {\omega_{k}^{2} \over k_{\parallel} \vert k_{\parallel} \vert} {\left| {{e(\phi-\psi)_{k}} \over {T_{e}}} \right|^{2}} f_{o} {\mid_{\omega / k_{\parallel}}}\label{eq4c}.
\end{equation}
\end{mathletters}

\noindent
Here $\psi = \omega \widetilde{A}_{\parallel} / {ck_{z}}$ and $\phi$ and $\psi$ are related by $\psi_{k} = (1 + k_{\perp}^{2} \rho_{s}^{2})\phi_{k}$.  Only the non-adiabatic piece of $\widetilde{f}$ contributes to $\langle \widetilde{E}_{z}\widetilde{n} / n_{o} \rangle$ and $\Gamma$.  Note that the flux of parallel current is proportional to a spectrally averaged factor of $({\underline{k}_{\bot}} \times {\widehat{z})} / {k_{z}}$, which requires that the turbulence have a net spectral chirality, in order for $\underline{\Gamma}$ to be non-zero.   This property is the manifestation of helicity in the KSAW dynamo problem.  Note also that $\Gamma$ is independent of resistivity $\eta$, since KSAW Landau damping now provides the irreversibility which permits `fast' transport of current!

The stability of the KSAW spectrum to a zonal magnetic field perturbation $\underline{\delta B_{\theta}} = \delta B_{\theta} (r) \widehat{\theta}$ may be determined by modulating Eqn. (\ref{eq3}), i.e.

\begin{mathletters}
\begin{equation}
{1 \over c}{\partial \over \partial t} {\delta {\langle A \rangle}_{q}} - {{\delta \langle \widetilde{E}_{z}\widetilde{n} / n_{o} \rangle} \over {\delta \langle A \rangle}} {\delta {\langle A \rangle}_{q}} + iq {{\delta\Gamma_{r}} \over {\delta \langle A \rangle}} {\delta {\langle A \rangle}_{q}} = {{-\eta} q^{2} \over c} \delta {\langle A \rangle}_{q}\label{eq5a},
\end{equation}

\noindent
where $\delta{\langle A \rangle}$ is the associated modulation in the vector potential, i.e. $\delta B_{\theta}(r) = - \partial \delta {\langle A \rangle} / \partial r$.  Anticipating the use of methods from adiabatic theory, $\langle {\widetilde{E}_{z}}\widetilde{n} / n_{o} \rangle$ and $\Gamma_{r}$ are conveniently re-expressed in terms of the KSAW action density $N_{k}$ as

\begin{equation}
{\langle{\widetilde{E}_{\parallel}} {\widetilde{n} \over n_{o}} \rangle} = - {\pi \over e} T_{e} \sum_{\underline{k}} {k_{\parallel}^{2}} {{2 \rho_{s}^{2} k_{\perp}^{2}} \over {2 + \rho_{s}^{2} k_{\perp}^{2}}} {{{\omega_{\underline{k}}^{2}} \over {k_{\parallel} {\vert  k_{\parallel} \vert}}}} f_{o} {\mid_{\omega / k_{z}}} {N_{\underline{k}}}\label{eq5b},
\end{equation}

\begin{equation}
{\Gamma_{r}} = {\pi \over e \Omega_{e}} T_{e} \sum_{\underline{k}} {{2 \rho_{s}^{2} {k_{\perp}^{2}} \over {2 + \rho_{s}^{2} k_{\perp}^{2}}}} k_{\theta} {{\omega_{\underline{k}}^{3}} \over {k_{\parallel} \vert k_{\parallel} \vert}} f_{o} {\mid_{\omega / k_{z}} N_{\underline{k}}}\label{eq5c},
\end{equation}

\noindent
where $N_{\underline{k}} = W_{\underline{k}} / \omega_{\underline{k}}$ and the KSAW energy density (normalized to the thermal energy density $n_{o}T_{e}$) is

\begin{equation}
W_{k} = {1 \over 8\pi n_{o}T_{e}} {({k_{\perp}^{2}} {\vert A_{\underline{k}}\vert^{2}} + {c^{2} \over v_{A}^{2}} k_{\perp}^{2} {\vert \phi_{\underline{k}} \vert^{2}})}\label{eq5d}.
\end{equation}
 
Using the relation between potentials $\phi_{k}$ and $A_{\parallel k}$, this can be re-expressed as:

\begin{equation}
{N_{\underline{k}}} = {{2 + \rho_{s}^{2} k_{\perp}^{2}} \over {2 \omega_{\underline{k}}}} \rho_{s}^{2} k_{\perp}^{2} \left|{e \phi_{\underline{k}} \over T_{e}} \right|^{2}\label{eq5e}.
\end{equation}
\end{mathletters}

\noindent
Here $\rho_{s} = c_{s}/\Omega_{i}$ and $c_{s}^{2} = T_{e}/m_{i}$.  In this formulation, then, the modulations $\delta\langle \widetilde{E}_{z} \widetilde{n} / n_{o} \rangle$ and $\delta \Gamma$ can now be calculated simply by computing the modulational response $\delta N_{\underline{k}} / \delta \langle A \rangle$.

The modulation in the number of KSAWs induced by the dynamo generated zonal field $\delta \langle A \rangle$, i.e. $\delta N_{\underline{k}} / \delta \langle A \rangle$, may be obtained by using the wave kinetic equation to determine the response $\delta N_{\underline{k}}$ to $\delta {\langle A \rangle}$.  The action density of a wave packet evolves according to the wave kinetic equation

\begin{equation}
{\partial N_{k} \over \partial t} + {\underline{v}_{g}} \cdot {{\underline{\nabla}}N_{k}}- {\underline{\nabla}} \omega_{k} \cdot {\partial N_{k} \over \partial {\underline{k}}} = - \gamma_{k} N_{k}\label{eq6},
\end{equation}

\noindent
where $\gamma_{k}$ is the wave damping decrement.  The dynamo generated field refracts the underlying KSAWs and thus induces spectral modulations.  In the presence of dynamo-generated zonal fields $\omega_{k} \rightarrow {\omega_{k}^{o}} + {\delta\omega_{k}}$.  Here $\omega_{k}^{2} = k_{\parallel}^{2}v_{A}^{2}(1+k_{\perp}^{2}\rho_{s}^{2})$ is the linear wave frequency and

\begin{equation}
\delta\omega_{\underline{k}} = {{\omega_{k}} \over {k_{z}B_{o}}} k_{\theta} {\delta B_{\theta}}\label{eq7},
\end{equation}

\noindent
is the perturbation in the wave frequency induced by the zonal dynamo field.  Linearizing Eqn. (\ref{eq6}) then gives the action density modulation

\begin{equation}
{\delta N_{k,q}} = {\left[{i \omega_{k}} \over {\Omega_{q} - q v_{g,r} + i \gamma_{k}}\right]} {{k_{\theta}} \over {k_{z}}} {{q^{2}\delta \langle A \rangle_{q}} \over {B_{o}}} {\partial \langle N_{k} \rangle \over \partial k_{r}}\label{eq8}.
\end{equation}

\noindent
Here $q$ is the radial wave number of the modulation $\delta{\langle A \rangle}$ and $\Omega_{q}$ is its frequency.  $\langle N_{k} \rangle$ is the mean KSAW action distribution.  Eqn. (\ref{eq8}) may then be substituted into Eqns. (\ref{eq5a},\ref{eq5b},\ref{eq5c}) to obtain the general expression for $\Omega_{q}$, which is (neglecting resistive dissipation)

\begin{mathletters}
\begin{equation}
{\Omega_{q} = {{\pi\rho_{e}^{2}d_{e}^{2}} v_{the}} q^{2} \sum_{\underline{k}}} {{2(1 + \rho_{s}^{2} k_{\perp}^{2})^{3 / 2}} \over {2 + \rho_{s}^{2} k_{\perp}^{2}}} {\Omega_{e} k_{z}^{2} + i \omega_{k} k_{\theta} q \over \Omega_{q} - q v_{g,r} + i \gamma_{k}} {k_{\perp}^{2} k_{\theta} \over \vert k_{\parallel} \vert} {\partial \over \partial k_{r}} {\left({<W_{k}>} \over {(1+\rho_{s}^{2} k_{\perp}^{2})^{1 / 2}}\right)}\overline{f}_{o} {\left({{\omega} \over {k_{\parallel}}}\right)}\label{eq9a}.
\end{equation}

\noindent
Here $d_{e} = c / \omega_{p,e}$, $\rho_{e}$ is the electron gyroradius and $f_{o} = \overline{f}_{o} / v_{the}$.  Now, noting that the radial group velocity is $v_{g,r} = {\partial\omega_{k} / \partial k_{r}} = ({k_{z}^{2}v_{A}^{2} / \omega_{k}}) k_{r} \rho_{s}^{2}$ and considering the limit $q v_{g,r} > \gamma$, $\Omega_{q}$ (note here $\Omega_{q}$ is small, i.e. of $O(\widetilde{E}_{\vert\vert}^{2})$, and $qv_{gr} > \gamma$ is consistent with the assumption of weak wave damping), we ultimately obtain the zonal field eigenfrequency $\Omega_{q}$, i.e.

\begin{eqnarray}
&&{{\Omega_{q}}=-{4\pi} {c_{s}}{d_{e}q}{\sum_{\underline{k}}}} {{(1+\rho_{s}^{2}k_{\perp}^{2})^{2}} \over {2 + \rho_{s}^{2} k_{\perp}^{2}}} k_{\theta}k_{\perp}^{2} {{k_{\parallel}} \over {\vert k_{\parallel} \vert}} {{\partial} \over {{\partial} {k_{r}^{2}}}} 
{\left({<W_{k}>} \over {(1+\rho_{s}^{2}k_{r}^{2})^{1/2}} \right)}\overline{f}_{o} {\left( {{\omega} \over {k_{\parallel}}} \right)}\nonumber\\
&&-i4{\pi}({c_{s}^{2} / v_{the})}{d_{e}^{2}}{q^{2}}{\sum_{\underline{k}}}{{(1+\rho_{s}^{2}k_{\perp}^{2})^{5/2}} \over {2 + \rho_{s}^{2} k_{\perp}^{2}}} {k_{\theta}^{2} {k_{\perp}^{2}} \over {\vert k_{\parallel}\vert}} {\partial \over {{\partial}{k_{r}^{2}}}} {\left({<W_{k}>} \over {(1+\rho_{s}^{2}k_{\perp}^{2})^{1/2}} \right)}\overline{f}_{o} {\left( {{\omega} \over {k_{\parallel}}} \right)}\label{eq9b}.
\end{eqnarray}
\end{mathletters}

Several aspects of this result are of interest.  First, note that zonal field growth requires a normal population profile, i.e. $\partial \langle N_{k} \rangle / \partial {k_{r}^{2}} < 0$, similarly to the case of zonal {\it flow} generation by drift waves.  This condition is virtually always satisfied for Alfvenic MHD turbulence.  Second, note that zonal field growth $Im\Omega_{q}$ is independent of $\eta$, since electron Landau damping provides the requisite dissipation.  However, concommitant with this is the fact that the dynamo drive is ultimately proportional to $\vert {\widetilde{E}}_{z\underline{k}}{\vert^{2}} / {k_{z}^{2}} = {\vert{({\phi}}-{{\psi}})_{\underline{k}}}{\vert^{2}}$, and thus depends directly upon field-fluid decoupling due to finite gyro-radius (and possibly finite ion inertial layer width, i.e. finite ${k_{z}^{2}}{c^{2}}/{\omega_{pi}^{2}}$) effects.  This is a direct consequence of the fact that finite $E_{\parallel}$ is required for Landau damping.  Since, in turn, Landau damping provides the necessary mechanism for irreversibility, the KSAW dynamo necessarily then requires non-zero $E_{\parallel}$ on KSAW scales. Note also that while a mean $\Gamma$ (and thus a mean field KSAW dynamo) requires a net chirality or helicity in the KSAW turbulence, {\it zonal field growth does not}.  This is because zonal field formation is a spontaneous symmetry breaking phenomenon, whereby the system acts to reinforce an initial $\delta {B_{\theta}}$ of either sign, but does {\it not} amplify the total flux.  In this respect, the zonal field amplification process is more like the small-scale dynamo than the mean field dynamo.  However, we hasten to add that a broad spectrum of growing zonal fields, with $(\rho_{i} / L_{\perp}) < q {\rho_{i}} < 1$, can be expected (N.B. Here $L_{\perp}$ is the scale of the Alfven wave spectrum inhomogeneity in the direction perpendicular to $B_{o}{\widehat{z}}$).  Thus, the zonal field dynamo field is not restricted to hyper-fine scales such as $q\rho_{i}\sim 1$, etc.  A further observation is that the zonal field dynamo mode are predicted to oscillate as well as grow, i.e. $\Omega_{q}$ is complex.  A net chirality (i.e. finite spectrum averaged $k_{\theta}k_{z}$) is required for Re $\Omega_{q} {\not =} 0$.

Since KSAWs (with finite $k_{r}$) will be refracted by zonal fields, the KSAW spectrum will necessarily be modified by the growth of the dynamo-generated fields.  Thus, a self-consisted KSAW dynamo theory, which treats both the zonal field generation as well as their back-reaction on the KSAW's, is called for.  The effects of random refraction of KSAW's by the generated fields can be described by the quasilinear wave kinetic equation

\begin{mathletters}
\begin{equation}
{{\partial \langle N_{k} \rangle} \over {\partial t}} - {\partial \over \partial k_{r}} {\langle {{\partial \widetilde{\omega}_{k}} \over {\partial x}} \widetilde{N}_{k} \rangle} = -\gamma_{\underline{k}} \langle N_{k} \rangle\label{eq10a},
\end{equation}

which, using $\widetilde{\omega}_{k}$ and $\widetilde{N}_{k}$ as given by Eqns. (\ref{eq7},\ref{eq8}), can be simplified to

\begin{equation}
{\partial \langle N_{k} \rangle \over \partial t} - {\partial \over \partial k_{r}} D_{k} {\partial \langle N_{k} \rangle \over \partial k_{r}} = - \gamma_{\underline{k}} \langle N_{k} \rangle \label{eq10b},
\end{equation}

where:

\begin{equation}
D_{k} = {({\omega_{k} k_\theta \over k_{z}B_{o}})^{2}} \sum_{q} q^{2} {\vert \delta B_{\theta, q} \vert^{2}} R(\Omega_{q}-qv_{gr})\label{eq10c},
\end{equation}

\begin{equation}
R(\Omega_{q} - q v_{gr}) = \gamma_{\underline{k}} / \lbrack (\Omega_{q}-q v_{gr})^{2} + \gamma_{\underline{k}}^{2} \rbrack \cong {\gamma_{\underline{k}} / {(q v_{gr})^{2}}}\label{eq10d}.
\end{equation}
\end{mathletters}

\noindent
Here, the irreversibility intrinsic to $D_{k}$ is provided by KSAW electron Landau damping via $\gamma_{\underline{k}}$.  Since zonal field frequencies are low (i.e. $\Omega_{q} \sim (\delta B / B_{o})^{2}$), $D_{k}$ is effectively non-resonant in character.  Note the effect of random refraction is to 'diffuse' $k_{r}$ to higher values.  This refraction occurs as KSAW packets traverse the layered structures of zonal fields.  Since wave dissipation is likely to be stronger at high $k_{r}$, this constitutes a route for feedback of the fields on the KSAW intensity, as well as its spectral distribution. 

Having obtained the evolution equation for $\langle N_{k} \rangle$, we can now self-consistently describe the coupled evolution of the dynamo driven zonal field spectrum and the KSAWs.  These may be described by coupled predator-prey type equations for $\vert \delta B_{\theta, q} \vert^{2}$ and $\langle N_{k} \rangle$, which are:

\begin{mathletters}
\begin{equation}
{\partial \over \partial t} {\vert \delta B_{\theta, q} \vert^{2}} = 2q^{2} (\overline{\gamma}_{q} - \eta)\label{eq11a},
\end{equation}

and

\begin{equation}
{{\partial \langle N_{k} \rangle} \over {\partial t}} = {\partial \over \partial k_{r}} D_{k} {{\partial \langle N \rangle} \over {\partial k_{r}}} - \gamma \langle N_{k} \rangle\label{eq11b},
\end{equation}

where

\begin{equation}
\overline{\gamma}_{q} = - 4 \pi (c_{s}^{2} d_{e}^{2}/v_{the}) {\sum_{\underline{k}}} {{(1 + k_{\perp}^{2} \rho_{s}^{2})^{5 / 2}} \over {(2 + k_{\perp}^{2} \rho_{s}^{2})}} {{k_{\theta}^{2} k_{\perp}^{2}} \over {\vert k_{\parallel}\vert}} {\partial \over \partial k_{r}^{2}} {\left({<W_{\underline{k}}>} \over {(1 + k_{\perp}^{2} \rho_{s}^{2})^{1 / 2}} \right)} \overline{f}_{o} {\left( {{\omega} \over {k_{\parallel}}} \right)}\label{eq11c}.
\end{equation}
\end{mathletters}

\noindent
Here $\overline{\gamma}_{q} = \overline{\gamma}_{q} (\langle N \rangle)$ is derived directly from Eqn. (\ref{eq9b}) for $Im(\Omega_{q})$.  Note that Eqn. (\ref{eq11a}) implies that a critical KSAW intensity level $\langle N \rangle_{crit} \sim \eta$ is required for zonal flow growth via modulational instability.  For weak $\eta$, this implies that only a very modest level of KSAW excitation is required for the dynamo.  Moreover, in regimes of weak dissipation, it suggests that the `marginal point' for the KSAW dynamo will scale with $\eta$, despite the fact that $\eta$ is otherwise completely irrelevant to the dynamics of zonal field generation.  Eqn. (\ref{eq11b}) suggests that as zonal fields grow, high $k_{r}$ components in the KSAW spectrum will be generated, thus ultimately quenching KSAW energy via coupling to dissipation.  Finally, it should be noted that the expression for $\overline{\gamma}_{q}$ given in Eqn. (\ref{eq11c}) assumes $q\rho_{i} < 1$.  For $q\rho_{i} \sim 1$, additional FLR factors enter which force  $\overline{\gamma}_{q}$ to decay for $q\rho_{i} \geq 1$.

A moment's consideration of Eqns. (\ref{eq11a},\ref{eq11b},{\ref{eq11c}) naturally begs the question of what happens in the limit of $\eta \rightarrow 0$ since, in this case, there is no apparent control on zonal flow growth.  We speculate here that as $\eta \rightarrow 0$, zonal fields may be subject to collisionless reconnection instabilities (i.e. tearing modes), which limit zonal field growth without requiring resistive dissipation.  Such instabilities are analogous to Kelvin-Helmholtz type instabilities which may limit zonal flow growth in the zero collisionality limit \cite{ref9}.  Alternatively, the zonal fields may regulate themselves via feedback on $\partial \langle N_{k} \rangle / \partial k_{r}^{2}$ (i.e. by modifying the wave spectrum \cite{ref10}) or by trapping of KSAW packets \cite{ref11}.  The detailed dynamics of these collisionless reconnection instabilities and of the system's behavior for $\eta \rightarrow 0$ will be discussed in detail in a future publication.

In conclusion, we have demonstrated that a zonal field dynamo can be driven by collisionless kinetic Alfven wave turbulence.  This dynamo is `fast' in that the rate of requisite magnetic reconnection is determined by electron Landau damping and thus is independent of the collisional resistivity.  Concommitant with this, non-zero $E_{\vert\vert}$ (i.e. as due to FLR or finite ion inertial layer width) is required for magnetic field amplification.  The zonal field dynamo self-regulates via refraction-induced diffusion of the KSAW spectrum toward high $k_{r}$.  These predictions should be amenable to investigation in laboratory experiments.  Finally, it is interesting to note that for typical interstellar medium parameters $(T_{e} \raisebox{-1ex}{\~} 1 ev, B_{o} \raisebox{-1ex}{\~} 10^{-6}$ G, L $\raisebox{-1ex}{\~}$ 1 pc., etc.) and taking $B / B_{o} \raisebox{-1ex}{\~} 10^{-3}$, Eqn. (\ref{eq9b}) predicts that the magnetic field grows at a rate $\gamma_{o} \raisebox{-1ex}{\~} 10^{-17} sec^{-1}$, consistent with that needed to achieve equipartition in $10^{9}$ years.

We thank L. Chen, G. Falkovich, T.S. Hahm, E.-J. Kim and M. Malkov for stimulating discussions.  We also thank E.-J. Kim for a careful, critical reading of the manuscript.  This research was supported by Department of Energy Grant No. FG-03-88ER53275.  P.H. Diamond also acknowledges partial support from the National Science Foundation under Grant No. PHY99-07949, to the Institute for Theoretical Physics at U.C.S.B., where part of this work was performed.

\end{document}